\documentclass[a4paper,11pt]{article}

\usepackage{adjustbox}
\usepackage{amsfonts}
\usepackage{amssymb, amscd}
\usepackage{graphicx}
\usepackage{empheq}
\usepackage{mathrsfs}
\usepackage[table,xcdraw]{xcolor}
\usepackage{colortbl}
\usepackage{slashed}
\usepackage{pdfpages}
\usepackage{dsfont}
\usepackage{braket}
\usepackage{sidecap}
\usepackage{comment}
\usepackage{arydshln}
\usepackage[font=small,labelsep=none]{caption}
\usepackage{mathtools}
\usepackage{cite}
\usepackage{hyperref}
\usepackage{tikz}
\usetikzlibrary{arrows,decorations.markings}
\usepackage{subcaption}
\usepackage[mode=buildnew]{standalone}
\usepackage[normalem]{ulem}
\usepackage[left=2cm,right=2cm,top=3.5cm,bottom=3.5cm]{geometry}

\hypersetup{
    colorlinks=true,
    linkcolor=ceruleanblue,
    filecolor=ceruleanblue,      
    urlcolor=ceruleanblue,
    citecolor=ceruleanblue,
}

\definecolor{myblue}{RGB}{174, 198, 219}
\definecolor{myred}{RGB}{157,31,68}
\definecolor{grey}{rgb}{0.4,0.4,0.4}
\definecolor{dullmagenta}{rgb}{0.4,0,0.4}
\definecolor{darkblue}{rgb}{0,0,0.4}
\definecolor{midblue}{rgb}{0,0,0.5}
\definecolor{midred}{rgb}{0.5,0,0}
\definecolor{orange}{rgb}{1,0.5,0}
\definecolor{lightbrown}{rgb}{0.75,0.5,0.25}
\definecolor{tan}{cmyk}{0.14,0.42,0.56,0}
\definecolor{djunglegreen}{cmyk}{0.99,0,0.52,0}
\definecolor{lightgreen}{rgb}{0,1,0}
\definecolor{olivegreen}{cmyk}{0.64,0,0.95,0.40}
\definecolor{midgreen}{rgb}{0.0,0.675,0.0}
\definecolor{darkgreen}{rgb}{0,0.5,0}
\definecolor{ceruleanblue}{rgb}{0.0, 0.2, 0.7}
\definecolor{burgundy}{rgb}{0.5, 0.0, 0.13}
\definecolor{blue_light}{RGB}{0, 102, 204}

\newlength\mytemplen
\newsavebox\mytempbox

\makeatletter
\newcommand\mybox{%
    \@ifnextchar[
       {\@mybox}%
       {\@mybox[0pt]}}

\def\@mybox[#1]{%
    \@ifnextchar[
       {\@@mybox[#1]}%
       {\@@mybox[#1][0pt]}}

\def\@@mybox[#1][#2]#3{
    \sbox\mytempbox{#3}%
    \mytemplen\ht\mytempbox
    \advance\mytemplen #1\relax
    \ht\mytempbox\mytemplen
    \mytemplen\dp\mytempbox
    \advance\mytemplen #2\relax
    \dp\mytempbox\mytemplen
    \colorbox{black!10!white}{\hspace{1em}\usebox{\mytempbox}\hspace{1em}}}

\def\MP{M_{\rm Pl}}
\newcommand*{\Cc}{\mathcal}%
\newcommand{\vect}[1]{\boldsymbol{#1}}
\newcommand{\de}{\, \mathrm{d}}	 

\def\b{\tilde{b}}
\def\ii{{\rm i}}

\newcommand{\ex}[1]{{\rm e}^{#1}}

\newcommand{\be}{\begin{equation}}      

\numberwithin{equation}{section}

\tikzset{cross/.style={cross out, draw=black, minimum size=2*(#1-\pgflinewidth), inner sep=0pt, outer sep=0pt},
cross/.default={4pt}}

\tikzset{
    master/.style={
        execute at end picture={
            \coordinate (lower right) at (current bounding box.south east);
            \coordinate (upper left) at (current bounding box.north west);
        }
    },
    slave/.style={
        execute at end picture={
            \pgfresetboundingbox
            \path (upper left) rectangle (lower right);
        }
    }
}

\date{\today}

\begin{document}
	\begin{flushright} {\footnotesize IPMU25-0042}  \end{flushright}
	
	\begin{center}
        \LARGE{\bf Large $n$-point Functions in Resonant Inflation}
		\\[1cm] 
		
		\large{Paolo Creminelli$^{\,\rm a, \rm b, \rm c}$, S\'ebastien Renaux-Petel$^{\,\rm d}$, Giovanni Tambalo$^{\,\rm e}$ \\ and Vicharit Yingcharoenrat$^{\,\rm f, \rm g}$}
		\\[0.5cm]
		
		\small{
			\textit{$^{\rm a}$
				ICTP, International Centre for Theoretical Physics\\ Strada Costiera 11, 34151, Trieste, Italy}}
		\vspace{.2cm}
		
		\small{
			\textit{$^{\rm b}$
				IFPU - Institute for Fundamental Physics of the Universe,\\ Via Beirut 2, 34014, Trieste, Italy}}
		\vspace{.2cm}

        \small{
			\textit{$^{\rm c}$
				INFN, Sezione di Trieste, via Valerio 2, 34127 Trieste, Italy}}
		\vspace{.2cm}

		\small{
			\textit{$^{\rm d}$
				Institut d'astrophysique de Paris, UMR 7095 du CNRS et de Sorbonne Universit\'e,\\ 98 bis bd Arago, 75014 Paris, France}}
		\vspace{.2cm}
		
		\small{
			\textit{$^{\rm e}$
				Institut f\"ur Theoretische Physik, ETH Z\"urich, Wolfgang-Pauli-Str. 27, 8093 Z\"urich, Switzerland}}
		\vspace{.2cm}

            \small{
            \textit{$^{\rm f}$
                High Energy Physics Research Unit, Department of Physics, Faculty of Science, Chulalongkorn University, Pathumwan, Bangkok 10330, Thailand}}
            \vspace{.2cm}
            
		\small{
			\textit{$^{\rm g}$
				Kavli Institute for the Physics and Mathematics of the Universe (WPI), The University of Tokyo Institutes for Advanced Study (UTIAS), The University of Tokyo, Kashiwa, Chiba 277-8583, Japan}}
		\vspace{.2cm}

	\end{center}
	
	\vspace{0.3cm} 
	
	\begin{abstract}\normalsize
    We investigate a qualitatively new regime of inflationary models with small and rapid oscillations in the potential---resonant non-Gaussianity. In contrast to the standard scenario, where most of the observable information is encoded in the power spectrum, in this regime the oscillatory signal predominantly appears in higher-order correlation functions with large $n$. This behavior emerges when the oscillation frequency $\omega$ exceeds the naive cutoff of the theory, $4\pi f$. However, as noted by Hook and Rattazzi \cite{Hook:2023pba}, the actual cutoff is somewhat higher---though only logarithmically---when the amplitude of the oscillations is small. We identify a phenomenologically relevant window in which $n$-point functions with $3 \lesssim n \lesssim 9$ are potentially observable. In this regime, the signal exhibits 350 -- 1000 oscillations per decade in $k$.

	\end{abstract}
	
	\vspace{0.3cm} 
	
	\vspace{2cm}
	
	\newpage
	{
		\hypersetup{linkcolor=black}
		\tableofcontents
	}
	
	\flushbottom
	
	\vspace{1cm}

\section{Introduction and setup}\label{sec:intro}

It is generally believed that most of the information of inflationary correlators is contained in $n$-point functions with small $n$.
For this reason, one typically focuses on the power spectrum and a limited number of higher-order correlation functions, such as the bispectrum and trispectrum.
Specifically, in a single-field model of inflation with oscillatory potential, it was shown in \cite{Behbahani:2011it} that within the regime of validity of the  effective field theory (EFT), the oscillatory signal contained in the $2$-point function dominates over the signal contained in higher-order $n$-point functions.
Note that $n$-point correlation functions of the curvature perturbation have been extensively studied in the literature, see e.g.~\cite{Chen:2008wn,Leblond:2010yq, Flauger:2010ja}.
However, the analysis done in \cite{Behbahani:2011it} relied on the fact that the EFT is expected to break down at the \textit{naive} UV cutoff $4 \pi f$, beyond which the theory becomes strongly coupled.
This implies that if one goes beyond such a scale, there might be phenomenologically interesting signals one has not encountered before. 
Recently in \cite{Hook:2023pba}, the perturbative unitarity bound for a canonical scalar-field model with oscillatory potential was reconsidered, revealing that the UV cutoff of the EFT exceeds the scale $4\pi f$. 
Therefore, this opens up a window in which the oscillatory signal encoded in higher-order $n$-point functions may become potentially observable and dominate that of the $2$-point function.
Scrutinizing this is the main goal of this paper.

In the resonant model, we assume that a canonical scalar field $\phi$ drives inflation with the potential
\begin{align}\label{eq:potential}
    V(\phi) = V_{\rm sr}(\phi) + \Lambda^4 \cos(\phi/f) \;,
\end{align}
where $V_{\rm sr}(\phi)$ is a slow-roll potential, $\Lambda$ is the scale that controls the amplitude of the oscillatory potential and $f$ is the energy scale analogous to the axion decay constant.   
The detailed analysis of axion monodromy inflation with $V_{\rm sr}(\phi) = \mu^3 \phi$ can be found in \cite{Flauger:2009ab}. 
Here we keep the slow-roll potential generic, as it was done in \cite{Flauger:2010ja,Leblond:2010yq}. During inflation the background metric is assumed to be the flat Friedmann-Lema{\^i}tre-Robertson-Walker (FLRW) metric: ${\rm d}s^2 = -\de t^2 + a(t)^2 \de \vect{x}^2$ where $a(t)$ is the scale factor.
Following the analysis in \cite{Leblond:2010yq,Flauger:2010ja,Flauger:2009ab} (see also Sec.~3.1 of \cite{Creminelli:2024cge}), the solution for the Hubble parameter $H(t) \equiv \dot{a}/a$, where a dot denotes a derivative with respect to $t$, up to first order in the oscillatory amplitude can be obtained analytically as
 \begin{align}
    \dot{H}(t) = -\epsilon_0 H_0^2 \big[1 - \tilde{b}\cos(\phi_0/f)\big] \;, \label{eq:hubble}
\end{align}   
where $0$ denotes quantities at zeroth order in the oscillatory component, and we have defined 
\begin{align}\label{eq:epsilon_tilde_b}
\epsilon_0 \equiv -\frac{\dot{H}_0}{H_0^2} = \frac{\dot{\phi}_0^2}{2\MP^2 H_0^2} \;, \qquad \tilde{b} \equiv \frac{2\Lambda^4}{\dot{\phi}_0^2} \;.
\end{align}
Note that it is straightforward to analytically obtain the solution for $\phi(t)$ up to first order in $\tilde{b}$ using Eq.~\eqref{eq:hubble} and $-2\MP^2 \dot{H} = \dot{\phi}^2$.
In this paper, we are interested in the regime where the frequency of oscillations is large compared to the Hubble rate and the oscillation amplitude is small, i.e.
\begin{align}
\alpha \equiv \frac{|\dot{\phi}_0|}{H_0f} = \frac{\omega}{H_0} \gg 1 \;, \qquad \tilde{b} \ll 1 \;, 
\end{align}
where $\omega \equiv |\dot{\phi}_0|/f$. This is indeed the regime where the $n$-point correlation functions of curvature perturbation were previously computed in \cite{Leblond:2010yq,Behbahani:2011it}.

Having discussed the dynamics of the background, we now focus on the analysis of perturbations around it. 
In this paper, for simplicity, we are interested in scalar fluctuations in the decoupling limit where the slow-roll parameter $\epsilon$ is taken to be very small. 
This implies that during inflation the metric becomes exactly de Sitter, for which the scale factor is $a(\eta) = -1/(H\eta)$ and $\eta$ is the conformal time with $\eta \in (-\infty,0]$.
In this limit, one can of course study the dynamics of inflaton perturbations using the action of a canonical scalar field with the potential \eqref{eq:potential}. It is also quite convenient, as pointed out in \cite{Creminelli:2024cge} (see also \cite{Firouzjahi:2025ihn}), to employ the EFT of inflation \cite{Cheung:2007st}. 
Without entering into the detailed derivation, it was shown that in the decoupling limit the action for $\pi(t,\vect{x})$ (the Goldstone boson that non-linearly realises the time diffeomorphism) is given by 
\begin{align}\label{eq:action_pi}
    S[\pi] = M_\textrm{Pl}^2\int \de^4x \, a(t)^3 \, \dot{H}(t+\pi)(\partial_\mu \pi)^2 \;, 
\end{align}
where higher-order terms in $\epsilon$ have been neglected. 
Note that the above form of the action we are considering is equivalent to that used to derive $n$-point functions in \cite{Behbahani:2011it,Leblond:2010yq}. 
This compact form of the action \eqref{eq:action_pi} encapsulates all nonlinearities/features of any model of canonical single-field inflation in the decoupling limit.\footnote{Although we are not interested here in the large-field limit (corresponding to the tail of the distribution function), it is useful to point out that the action \eqref{eq:action_pi} can be used to compute the wavefunction of the Universe in the large $\pi$ limit, as shown in \cite{Creminelli:2024cge}.}
For example, in the resonant model where the Hubble scale is given by Eq.~\eqref{eq:hubble} one sees that all nonlinearities of $\pi$ are encoded inside the cosine function.
Nevertheless, in the present paper, our calculation relies on standard perturbation theory, i.e.~we are not interested in non-typical fluctuations.  
Therefore, from the action \eqref{eq:action_pi} one can straightforwardly compute $n$-point correlation functions of $\pi$ using the well known \textit{in}-\textit{in} formalism \cite{Maldacena:2002vr,Weinberg:2005vy}.

As noted before, the naive UV cutoff of the model is at the scale $4 \pi f$.
However, a recent analysis in \cite{Hook:2023pba} showed that this EFT possesses a unitarity cutoff that lies above this estimate.\footnote{See also \cite{Craig:2019zkf,Ekhterachian:2021rkx} for other models whose unitarity cutoff was found to be higher than the usual scale $4\pi f$.}
In Sec.~\ref{sec:cutoff} we will reproduce their result in detail.  
Remarkably, as we will see, the fact that the revised unitarity cutoff exceeds $4 \pi f$ by logarithmic corrections opens a phenomenologically interesting window where large $n$-point functions are potentially measurable.  
The computation of the signal-to-noise ($S/N$) ratio based on the revised cutoff will be carried out in Sec.~\ref{sec:SNR}. 
In Sec.~\ref{sec:result} we present a range of parameters in which large $n$-point functions can be detectable, see in particular fig.~\ref{fig:snr_contour}, and discuss possible constraints in the high-frequency regime. 
Finally, we draw our conclusions and highlight potential directions for future work in Sec.~\ref{sec:conclusions}.

\section{Calculation of the unitarity cutoff}\label{sec:cutoff}

In this section we revisit the derivation of the unitarity cutoff derived in \cite{Hook:2023pba}. In the single-field model with the oscillatory potential $V(\phi) = \Lambda^4 \cos(\phi/f)$, the $n \to n$ scattering amplitude $\mathcal M_{n \to n}$ is given by
\begin{align}\label{eq:amplitude_E_dep}
	\mathcal M_{n\rightarrow n}  
	= 
	\frac{\lambda_{2n} }{8\pi} 
	\bigg(\frac{E}{4\pi}\bigg)^{2n-4} 
	\frac{1}{n!(n-1)! (n-2)!}  
	\;,
\end{align}
where $\lambda_{2n} = (-1)^n \Lambda^4 / f^{2 n}$ and $E$ is the center-of-mass energy.
In App.~\ref{app:flat_scattering} we give a detailed derivation of Eq.~\eqref{eq:amplitude_E_dep} (see also \cite{Chang:2019vez}). 
Written in terms of the resonant-model parameters ($\tilde{b}$ and $\omega$), Eq.~\eqref{eq:amplitude_E_dep} becomes 
\begin{align}\label{eq:M_n_to_n}
    \mathcal M_{n\rightarrow n}  
    =
    (-1)^n \tilde{b} \pi \bigg(\frac{\omega}{4\pi f}\bigg)^{2n-2}\frac{1}{n!(n-1)! (n-2)!} \;,
\end{align}
where we used $\tilde{b} = 2\Lambda^4/(\omega^2 f^2)$ and we have identified the energy $E$ with the oscillation frequency $\omega$. Indeed, as we will see, the total energy of the perturbations in a given $n$-point function is given by $\omega$.\footnote{Note that since the scattering amplitude \eqref{eq:amplitude_E_dep} with energy $E$ is an object defined on flat space, identifying $E$ with $\omega$ (oscillation frequency) is valid when the mode functions are inside the horizon. This is consistent with the fact that the saddle point at which $n$-point functions of curvature perturbation are peaked is at $-k_{\rm t} \eta_{\rm s} \simeq \alpha \gg 1$, where $k_{\rm t}$ is the total momentum and $\eta_s$ is the saddle point. If $\eta_s$ moves to later times due to large $n$, one then expects corrections in the identification of $E$ and $\omega$.}
The perturbative unitarity bound is obtained by imposing $|\mathcal M_{n\rightarrow n}| < 1$.
The maximum of $|\mathcal M_{n\rightarrow n}|$ over $n$ is obtained by solving $\partial_n |\mathcal M_{n\rightarrow n}| = 0$. By expanding the latter equation for large $n$ we obtain that the maximum $n$ is attained at
\begin{equation}\label{eq:n_max_M}
	n_{\rm max}
	\simeq
	\left(
        \frac{\omega}{4 \pi f}
	\right)^{2/3}
	\;.
\end{equation}
Notice that we obtain a different prefactor compared to Eq.~(2.4) in \cite{Hook:2023pba}. We checked that our result gives a better approximation to the true maximum of the amplitude. 
From Eq.~\eqref{eq:n_max_M} it is clear that when $\omega / (4 \pi f)$ is large, $n_{\rm max}$ is also large. 
The condition $|\mathcal M_{n\rightarrow n}| < 1$  evaluated at $n_{\rm max}$ gives the maximum value for $\omega < \Lambda_{\rm cutoff}$. 
Therefore, in the large-$n$ limit, we obtain the implicit expression
\begin{equation}\label{eq:cutoff_resonant}
	\frac{\Lambda_{\rm cutoff}}{4 \pi f}
    \simeq \bigg\{
		\frac{1}{3}
        \log 
		\Big[
			\frac{2}{\b}\Big(\frac{f}{\Lambda_{\rm cutoff}}\Big)^{2}
		\Big]
	\bigg\}^{3/2}
	\;,
\end{equation}
where we have dropped terms that are not enhanced by the logarithmic factor. 
From the expression \eqref{eq:cutoff_resonant}, one sees that $\Lambda_{\rm cutoff}$ is larger than the naive cutoff $4 \pi f$ by a logarithmic factor that depends on $\tilde{b}$. 
Notice that, as $\tilde{b} \to 0$, this cutoff approaches infinity, i.e.~the theory becomes free as oscillations in the potential disappear. 
One can explore physics beyond the scale $4 \pi f$ up to $\Lambda_{\rm cutoff}$ using our EFT. 
However, for the values of $\b$ and $\omega$ relevant for our results, the approximation \eqref{eq:cutoff_resonant} does not suffice to have a good accuracy. For this reason in our results (Sec.~\ref{sec:result}) we instead evaluate the cutoff numerically: for a given $\b$ the allowed $\omega$'s have to satisfy $|\Cc M_{n \to n}| < 1$ for all integer $n \geq 2$.

In this paper we remain agnostic about the possible UV completions of the EFT and we only require to be below the unitarity cutoff. It is not guaranteed that a UV completion without extra states below $\Lambda_{\rm cutoff}$ exists. For example, the explicit partial UV completion of the effective model studied in \cite{Hook:2023pba} introduces states which are parametrically lighter than the $\Lambda_{\rm cutoff}$ derived above.\footnote{See also \cite{Hook:2019mrd} for a UV completion of the partial UV model in \cite{Hook:2023pba}. This UV theory involves $\ell$ vector-like fermions and a complex scalar field.} We proceed to compute, within the EFT, the $n$-point correlation functions and their corresponding signal-to-noise ratios, with the goal to identify the parameter space where such signals are detectable.  
It is important to mention that in the presence of additional degrees of freedom lighter than $\Lambda_\textrm{cutoff}$, arising from a (partial) UV completion, the phenomenology associated with the $(S/N)$ analysed in this paper should be revisited.


\section{Signal-to-noise ratio of the $n$-point functions}\label{sec:SNR}

The connected $n$-point correlation function of $\zeta$ can be analytically computed at first order in $\b$. Their explicit expressions for large $\alpha$ are \cite{Leblond:2010yq,Behbahani:2011it}
\begin{align}\label{eq:corre_res}
	\braket{\zeta_{\vect k_1} \cdots \zeta_{\vect k_n}}
	= (2\pi)^3 
    \delta^{(3)}(\vect k_{\rm t}) 
    A_n B_n(k_i) 
	\;,
\end{align}
where 
\begin{align}\label{eq:resonant_An}
    A_n 
    = 
    (-1)^{n+1} 
    \frac{\sqrt{2\pi} \, \b}{2^{n+1}} 
    \alpha^{2n-7/2} P_\zeta^{n-1} \;,
\end{align}
and
\begin{align}\label{eq:col_B_N}
	B_n(k_i) 
	= 
	\frac{1}{k_{\rm t}^{n-3} \prod_i k_i^2} 
	\bigg[ 
		\sin(\alpha\log(k_{\rm t}/k_\star)) 
		+ \frac{1}{\alpha}  \cos(\alpha\log(k_{\rm t}/k_\star))  
		\sum_{i,j} \frac{k_i}{k_j}
	\bigg] 
	\;,
\end{align}
with $\vect k_{\rm t} \equiv \sum_i \vect k_i$, $ k_{\rm t} \equiv \sum_i k_i$, $P_\zeta = H^4/(2\MP^2 |\dot{H}_0|)$ and $k_\star$ is a pivot scale. 
For $n=2$, notice that the power spectrum also has a smooth component besides this resonant signal. The second term in the square bracket of $B_n$, suppressed by $1 / \alpha$, will be neglected in the following.\footnote{In this paper, we do not take into account the squeezed limit contribution of Eq.~(\ref{eq:col_B_N}), which is encoded in subleading terms in $\alpha$. However, taking these terms into account, it was shown in \cite{Kalaja:2020mkq} that the $(S/N)^2$ of $n$-point functions get a mild logarithmic enhancement $\sim \log(k_{\rm max}/k_{\rm min})$, which does not affect our conclusions.
} 
These $n$-point functions are the main ingredient we will use below in order to compute the signal-to-noise ratio.\footnote{
	  The oscillatory time dependence of the Hubble rate leads to a parametric resonance in the linear equations for $\pi$. Let us show that in the parameter space of interest this growth is not relevant. Neglecting momentarily the expansion of the universe, the linearised equation in momentum space for $\pi_{\vect k}(t)$ at linear order in $\b$ is $\ddot \pi_{\vect k} + (\vect k^2 / a^2) \pi_{\vect k} + \omega \b  \sin (\omega t) \dot \pi_{\vect k} = 0$. 
	  We can recast this into a Mathieu equation by defining $\varphi_{\vect k} \equiv (1 + \b / 2 \cos(\omega t)) \pi_{\vect k}$. Introducing $z \equiv \omega t / 2$, indeed we have $\varphi_{\vect k}'' + [A_{\vect k} + 2 q \cos (2 \tau)]\varphi_{\vect k} = 0$, where $A_{\vect k} \equiv 4 \vect k^2 / (\omega^2 a^2)$, $q \equiv - \b$ and $'$ here stands for a derivative with respect to $z$.

    For $q \ll 1$, the equation is in the narrow-resonance regime and it develops instability bands around $A_{\vect k} = \ell^2$ for $\ell$ integer. The largest growth is for $\ell = 1$, where the Floquet exponent is $\mu_{\vect k} \sim |q| / 2$ and the width of the band $A_{\vect k} \simeq 1 \pm |q|$.
    Then, the width translates into $k / a \simeq \omega  (1 \pm \b)/2$.
    Neglecting the expansion of the universe and evaluating the growth after one Hubble time $\Delta t \sim 2 \Delta z/\omega \sim H^{-1}$, we would obtain a growth $\exp(\mu_{\vect k} \Delta z) \sim \exp(\b \, \omega / (4H)) = \exp(\alpha \b/4)$. This is not quite correct since the momentum $\vect k$ redshifts and quickly exits the resonant band. This happens when $(\Delta k / a )/ (k / a) \sim H \Delta t \sim \b$, where the last requirement is the width of the band. Therefore, the accurate estimate for the growth is $\sim \exp(\alpha \b^2/4)$.
    We now argue that we are always in the regime where this growth is small. To see this, let us consider the correction to the power spectrum of $\zeta$ in the resonant model at order $\b$: it scales as $\sqrt{\alpha} \, \b$, see Eq.~\eqref{eq:corre_res}, and observationally this is bounded by $\sqrt{\alpha} \, \b \lesssim 10^{-2}$ \cite{Behbahani:2011it}.
    Therefore, we are always in the regime where $\exp(\alpha \b^2/4) \simeq 1$. Notice that the estimate for the growth depends on the same parameter as the correction to the power spectrum, as expected.
} (Note that the dimensionless power spectrum whose value is constrained to $2 \times 10^{-9}$ by CMB observations reads $\Delta_\zeta^2=P_\zeta/(4 \pi^2)$.)

These correlators are obtained by evaluating the \textit{in}-\textit{in} time integrals through a saddle-point approximation \cite{Leblond:2010yq}. More specifically, in terms of conformal time $\eta$, the saddle is located at $(-k_{\rm t} \eta_s) = \alpha$. Given that $\alpha$ is large, this can be understood as a resonance where the time-dependent interaction $\sim \cos(\omega t)$ produces $n$ quanta when their total frequency matches $\omega$. This implies that the total energy of the quanta is $\omega$, as we discussed above. The latter is then required to be below the perturbative unitarity cutoff found in the previous section. 
Indeed, unitarity requires new states lighter than $\Lambda_{\rm cutoff}$: if $\omega$ exceeds $\Lambda_{\rm cutoff}$ these new states would be generically produced invalidating the EFT calculation.


\subsection{Optimal estimator}
Given the $n$-point functions above, we would like to obtain the optimal estimator for the parameter $\b$.
To proceed in this direction, we first define appropriate estimators $\hat{\Cc E}_{n}$ for each $n$-point function, see e.g.~\cite{Munchmeyer:2019wlh}, as follows
\begin{equation}\label{eq:estimator_n}
	\hat{\Cc E}_{n}
	\equiv 
	\frac{1}{ N_n n!}
	\int \prod_{i = 1}^n 
	\bigg[\frac{\de^3 \vect{k}_i}{(2\pi)^3}\bigg] 
	(2 \pi)^{3}
	\delta^{(3)}(\vect k_{\rm t}) 
	\braket{ \zeta_{\vect{k}_1} \cdots \zeta_{\vect{k}_n}}'_{| \b=1 }
	\frac{\zeta_{\vect{k}_1} \cdots \zeta_{\vect{k}_n}}
	{P(k_1) \cdots P(k_n)} 
	\;.
\end{equation}
Here and in the following, $P(k) \equiv \braket{\zeta_{\vect k} \zeta_{-\vect k}}'=P_\zeta/(2 k^3)$ is the power spectrum and prime denotes the correlators without the factor $(2\pi)^3 \delta^{(3)}(\sum_i \vect{k}_i)$. 
The normalisation factor $N_n$, which we will compute later, is defined such that the estimator is unbiased, i.e.~$\langle \hat{\Cc E}_{n} \rangle=\b$. For simplicity we will assume that the integral over momenta in \eqref{eq:estimator_n} is restricted to non-degenerate polygons, i.e.~there is no subset of the wavevectors $\vect k_i$ whose sum vanishes. This allows us to neglect non-connected contributions and simplify the following algebra. We expect that this simplification does not affect the final results parametrically.\footnote{We are also neglecting terms in the optimal estimator that contain a lower number of $\zeta$'s \cite{Munchmeyer:2019wlh}.} 
The expectation values of these estimators read
\begin{equation}\label{eq:avg_estimator}
	\braket{\hat{\Cc E}_{n}}
	= 
	\frac{\b \, \Cc V}{N_n n!}
	\int \prod_{i = 1}^n 
	\bigg[\frac{\de^3 \vect{k}_i}{(2\pi)^3}\bigg] 
	(2 \pi)^{3}
	\delta^{(3)}(\vect k_{\rm t}) 
	\frac{[\braket{ \zeta_{\vect{k}_1} \cdots \zeta_{\vect{k}_n}}'_{| \b=1 }]^2}
	{P(k_1) \cdots P(k_n)} 
	\;,
\end{equation}
where we defined $\Cc V \equiv (2 \pi)^3 \delta^{(3)}(\vect 0)$ as the comoving volume (here the Dirac delta is in momentum space). This can be used to fix the value of $N_n$.

The optimal estimator is the linear combination of the individual estimators $\hat{\Cc E}_{n}$:
\begin{equation}
	\hat{\Cc E}_{\rm opt} 
	\equiv 
	\sum_{n \geq 2} \Cc W_n \hat{\Cc E}_{n}\,,
\end{equation}
with weights $\Cc W_n$ such that $\braket{\hat{\Cc E}_{\rm opt}} = \b$ and the estimator has the minimal variance $\braket{\hat{\Cc E}_{\rm opt}^2}-\braket{\hat{\Cc E}_{\rm opt}}^2$. The corresponding total signal-to-noise ratio squared is then naturally defined as $(S/N)^{2}_{\rm tot} \equiv \braket{\hat{\Cc E}_{\rm opt}}^2/ \braket{\hat{\Cc E}^2_{\rm opt}}$, and similarly for each individual estimator $\hat{\Cc E}_{n}$. By virtue of restricting estimators to non-degenerate configurations, off-diagonal terms in the covariance matrix $\braket{\hat{\Cc E}_{n} \hat{\Cc E}_{m}}$ vanish.\footnote{In App.~\ref{app:off-diagonal}, we show explicitly the effect of including degenerate configurations on the off-diagonal terms of the covariance matrix.}
Following the discussion in App.~A of \cite{Munchmeyer:2019wlh}, this makes the derivation of the optimal estimator much easier. First, notice that at the lowest order in $\b$, $\braket{\hat{\Cc E}^2_n}$ can be obtained by treating the fields $\zeta_{\vect{k_i}}$ in \eqref{eq:estimator_n} as Gaussian. One then obtains (again neglecting degenerate configurations)
\begin{align}\label{eq:var_estimator}
	\braket{\hat{\Cc E}_{n}^2 }
	&= 
	\frac{\Cc V}{N_n^2 n!}
	\int \prod_{i = 1}^n 
	\bigg[\frac{\de^3 \vect{k}_i}{(2\pi)^3}\bigg] 
	(2 \pi)^{3}
	\delta^{(3)}(\vect k_{\rm t}) 
	\frac{
		[\braket{ \zeta_{\vect{k}_1} \cdots \zeta_{\vect{k}_n}}'_{| \b=1}]^2}
	{P(k_1) \cdots P(k_n)}
	\;,
\end{align}
and the signals-to-noise in each $n$-point function:
\begin{align}
\label{individual-SNR}
\bigg(\frac{S}{N}\bigg)^2_{n}  \equiv \frac{\braket{\hat{\Cc E}_n}^2}{\braket{\hat{\Cc E}^2_n}}=\frac{\b^2 \Cc V}{n!}
	\int \prod_{i = 1}^n 
	\bigg[\frac{\de^3 \vect{k}_i}{(2\pi)^3}\bigg] 
	(2 \pi)^{3}
	\delta^{(3)}(\vect k_{\rm t}) 
	\frac{
		[\braket{ \zeta_{\vect{k}_1} \cdots \zeta_{\vect{k}_n}}'_{| \b=1}]^2}
	{P(k_1) \cdots P(k_n)}\,,
\end{align}
which are simply related to the normalisation factor as $(S/N)^{2}_n=\b^2 N_n$.

Now, minimizing $\braket{\hat{\Cc E}_{\rm opt}^2}=\sum_{n \geq 2} \Cc W_n^2 \braket{\hat{\Cc E}_{n}^2 }$ subject to $\sum_{n \geq 2} \Cc W_n=1$ can be straightforwardly done by introducing a Lagrange multiplier to enforce the constraint and taking the derivative with respect to $\Cc W_n$. This gives $\Cc W_n=N_n/(\sum_{m \geq 2} N_m)$, and the very intuitive result that the total signal-to-noise is the sum in quadrature of the signals-to-noise in each $n$-point function:
\begin{align}
\label{total-SNR}
\bigg(\frac{S}{N}\bigg)^2_{{\rm tot}}=\sum_n \bigg(\frac{S}{N}\bigg)^2_{n}\,,
\end{align}
where, from \eqref{eq:corre_res}, one explicitly has
\begin{align}\label{eq:SN_2}
	\bigg(\frac{S}{N}\bigg)^2_n 
	&=\Cc V 
    \,
	\frac{2^n A_n^2}{n! P_{\zeta}^n}
	\int \prod_{i = 1}^n 
	\bigg[
	\frac{\de^3 \vect{k}_i}{(2\pi)^3\,k_i}
	\bigg]
	(2\pi)^3 \delta^{(3)}
	\big(
	\vect k_{\rm t}
	\big)
	\, 
\frac{\sin^2(\alpha\log(k_{\rm t}/k_\star))}{k_{\rm t}^{2n-6}}  \;,
\end{align}
with $A_n$ given in \eqref{eq:resonant_An}.

\subsection{Evaluation of $(S/N)$}

Our aim is to evaluate the signal-to-noise ratio \eqref{eq:SN_2}.
First, we notice that in the limit of rapid oscillations $\alpha \gg 1$, we can replace the function $\sin^2$ by its average $1/2$. Indeed, writing $\sin^2 z = (1 - \cos 2 z)/2$, one sees that corrections are suppressed at large $\alpha$ due to oscillations in the momentum integrals $\sim \exp(-\alpha \pi)$.
The idea then is to factorize the $n$ integrals over momenta, in such a way that each integral can be computed separately. At first sight, the presence of $k_{\rm t}$ prevents us from proceeding. 
We therefore use a trick to replace $1/k_{\rm t}^{2n-6}$ with an integral over the Feynman variable $\tau$:
\begin{align}\label{eq:k_T_integral}
\frac{1}{k_{\rm t}^\beta} 
= 
\frac{1}{\Gamma(\beta)}
\int_{-\infty}^0 \de\tau 
\, 
(-\tau)^{\beta-1} \ex{k_{\rm t}\tau} 
\;,
\end{align}
with $\beta > 0$ and $k_{\rm t} > 0$.\footnote{For $n = 2$ and $n = 3$, one can straightforwardly evaluate the $(S/N)$ ratio without using the formula \eqref{eq:k_T_integral}.} We thus obtain
\begin{align}\label{eq:SN_3}
	\bigg(\frac{S}{N}\bigg)^2_n 
	= 
    \Cc V \, \frac{2^n  A_n^2}{2 \cdot n! P_{\zeta}^n}
    \frac{1}{\Gamma(2n- 6)} 
	\int \prod_{i = 1}^n 
	\bigg[
		\frac{\de^3 \vect{k}_i}{(2\pi)^3\,k_i}
	\bigg] 
	\int \de^3 \vect{x} \, 
	\ex{\ii \vect{x}\cdot\sum_i\vect{k}_i} 
	\int_{-\infty}^0 \de \tau \, 
	(-\tau)^{2n-7} \ex{k_{\rm t}\tau} 
 \;,
\end{align}
where we wrote 
$(2 \pi)^3\delta^{(3)}
\big(
\vect k_{\rm t}
\big) 
= \int \de^3 \vect{x}
\, 
\ex{\ii \vect{x} \cdot \sum_i \vect{k}_i}$. For a given experiment, the range of integration of momenta is limited by some cutoff $k_{\rm max}$: 
without this cutoff one would be measuring an infinite number of data and therefore the $(S/N)$ ratio would be divergent.
In particular, an experiment will have some characteristic window function going to zero above $k_{\rm max}$. In our calculation we assume an exponential window function of the form $\sim \exp(- k_i / k_{\rm max})$, which enters in each $\de^3 \vect k_{i}$ integral. With this choice, we are able  to perform analytical calculations of the $(S/N)$ ratio.\footnote{
We expect that a different choice of the window function will change the $(S/N)$ ratio by order unity, without affecting the conclusions.} Using the exponential window function, the expression \eqref{eq:SN_3} becomes
\begin{align}
	\bigg(\frac{S}{N}\bigg)^2_n 
	= 
    \Cc V \, \frac{ 2^n A_n^2}{2\cdot n! P_\zeta^n} 
    \frac{1}{\Gamma(2n- 6)} 
	\int 
	\prod_{i = 1}^n 
	\bigg[
	\frac{\de^3 \vect{k}_i}{(2\pi)^3\,k_i}
	\ex{- k_i / k_{\rm max}}
	\bigg] 
	\int \de^3 \vect{x} 
	\, 
	\ex{\ii \vect{x}\cdot\sum_i\vect{k}_i} 
	\int_{-\infty}^0 \de \tau \, 
    (-\tau)^{2n-7} \ex{k_{\rm t}\tau}  
	\;.
\end{align}
The integrals above are manifestly finite, so we are allowed to swap the order of the integrals $\de^3 \vect{k}_i$ with $\de \tau$ and $\de^3 \vect{x}$. 
Thus, we have 
\begin{align}
	\bigg(\frac{S}{N}\bigg)^2_n 
	& = 
	\Cc V  \,\frac{ 2^n A_n^2}{2\cdot n! P_\zeta^n}
    \frac{1}{\Gamma(2n- 6)}
	\int_{-\infty}^0 \de \tau 
	\, 
	(-\tau)^{2n-7}  
	\int \de^3 \vect{x} 
	\, 
	\prod_{i = 1}^n \int  
	\frac{\de^3 \vect{k}_i}{(2\pi)^3\,k_i}  
	\, 
	\ex{\ii \vect{x}\cdot \vect{k}_i} \ex{k_i\tau} \ex{-k_i/k_{\rm max}} 
	\nonumber \\ 
	& = 
	\Cc V
    \, \frac{ 2^n A_n^2}{2 \cdot n! P_{\zeta}^n} 
    \frac{1}{\Gamma(2n- 6)}
	\int_{-\infty}^0 \de \tau \, (-\tau)^{2n-7} 
	\int \de^3 \vect{x} 
	\bigg[
		\int \frac{\de^3 \vect{k}_i}{(2\pi)^3\,k_i}  
		\, 
		\ex{\ii \vect{x}\cdot \vect{k}_i} \ex{k_i\tau} 
		\ex{-k_i/k_{\rm max}} 
	\bigg]^n 
	\;.
	\label{eq:SNR_finite}
\end{align} 
The momentum integral over $\mathbb{R}^3$ in the last line of the above equation can be analytically computed as 
\begin{align}
    \int \frac{\de^3 \vect{k}_i}{(2\pi)^3\,k_i}  
		\, 
		\ex{\ii \vect{x}\cdot \vect{k}_i} \ex{k_i\tau} 
		\ex{-k_i/k_{\rm max}}  = 
        \frac{k_{\rm max}^2}{2\pi^2[k^2_{\rm max} r^2 + (1 - k_{\rm max} \tau)^2]} 
	\;, 
\end{align}
where we have defined $r = |\vect{x}|$.
Therefore, the $(S/N)$ ratio \eqref{eq:SNR_finite} becomes 
\begin{align}\label{eq:SNR_finite_2}
	\bigg(\frac{S}{N}\bigg)^2_n 
	= 
	\Cc V
    \, 
    \frac{ 2^n A_n^2}{2 \cdot n! P_\zeta^n}
    \frac{1}{\Gamma(2n- 6)}
	\int_{-\infty}^0 \de \tau 
	\, 
	(-\tau)^{2n-7} 
	\int \frac{\de^3 \vect{x} }{(2\pi^2)^n}
	\frac{ k_{\rm max}^{2n}}{[k^2_{\rm max} r^2 + (1 - k_{\rm max} \tau)^{2}]^n} \;. 
\end{align}
Then, performing the integral in $\vect{x}$ over $\mathbb{R}^3$ using spherical coordinates gives 
\begin{align}\label{eq:int_d3x}
	\int 
    \frac{\de^3 \vect{x} }
    {[k^2_{\rm max} r^2 + (1 - k_{\rm max} \tau)^{2}]^n}  
	= 
	\frac{\pi^{\frac{3}{2}} (1 - k_{\rm max} \tau)^{3 - 2n} 
    \Gamma(n - 3/2)
    }{k_{\rm max}^{3 } \Gamma(n)} 
	\;.
\end{align}
Using the above result in Eq.~\eqref{eq:SNR_finite_2}, we are thus left with a single integral over $\tau$, which can be computed analytically as well:
\begin{align}\label{eq:snr_int_over_t}
	\bigg(\frac{S}{N}\bigg)^2_n  
	& = 
	\frac{2^n \pi^{\frac{3}{2}} \Cc V  A_n^2}
	{2 (2\pi^2)^n n! P_\zeta^n} 
    \frac{k_{\rm max}^{2n - 3}}{\Gamma(2n- 6)}
	\frac{ 
    \Gamma(n - 3/2)
    }{\Gamma(n)}  
	\int_{-\infty}^0 \de \tau 
	\, 
	(-\tau)^{2n-7}  (1 - k_{\rm max} \tau)^{3 - 2n}  
	\nonumber \\ 
    & = 
    \frac{(4 \pi)^2 \Cc V A_n^2 }{(2\pi)^{2n} P_\zeta^n}
    \frac{k_{\rm max}^3}{n! (n-1)! (n-2)!}
	\;.
\end{align}
The factor $ \Cc V \  k_{\rm max}^3$ is related to the number of modes measured by an experiment. More precisely one can write
\begin{align}\label{eq:def_N_modes}
	\Cc V \ k_{\rm max}^3 
	=
	\frac{3 (2 \pi)^3 }{4\pi} 
     \Cc V 
	\int_0^{k_{\rm max}} 
	\frac{\de^3 \vect{k}_1}{(2\pi)^3} 
    = 
    \frac{3 (2 \pi)^3 }{4\pi}
    N_{\rm modes} 
	\;,
\end{align}
where $N_{\rm modes} \equiv \Cc V \int_0^{k_{\rm max}} \de^3 \vect{k}_1/(2\pi)^3 $ is the number of modes for an experiment with a sharp cutoff in Fourier space.
Therefore, Eq.~\eqref{eq:snr_int_over_t} gives 
\begin{align}\label{eq:SNR_finite_3}
	\bigg(\frac{S}{N}\bigg)^2_n 
	= 
	\frac
    {6 A_n^2}
    {(2\pi)^{2n - 4} P_\zeta^n}
    \frac{N_{\rm modes}}{n! (n-1)! (n-2)!}
	\;.
\end{align}
Using the expression \eqref{eq:resonant_An} for $A_n$ in Eq.~\eqref{eq:SNR_finite_3} we finally obtain 
\begin{empheq}[box={\mybox[5pt][5pt]}]{equation}
\label{eq:SNR_final}
	\bigg(\frac{S}{N}\bigg)^2_n  =
    \frac{3 \pi \alpha\, \b^2 }{16}
    \bigg(\frac{\omega}{4 \pi f}\bigg)^{2n - 4}
    \frac{1}{n! (n-1)! (n-2)!}
    N_{\rm modes}
	\;,
\end{empheq}
where we used $\alpha^2 P_\zeta^{1/2} = \omega / f$.
The expression above represents our exact result for the signal-to-noise ratio for any $n$, which we will use to evaluate the signal in each $n$-point function.\footnote{It is useful to trace the origin of the factorials appearing at the denominator. 
The first, $n!$, arises from the construction of the signal-to-noise ratio \eqref{individual-SNR}. The other two, $(n-1)!(n-2)!$, come from the phase-space integration over momenta. Indeed notice that one has the same factors in the S-matrix phase-space integral Eq.~\eqref{eq:omega_n}.}
\footnote{We checked that our result agrees with the ones for $n=2,3$ in \cite{Behbahani:2011it}, once correcting numerical factors there and modifying their computation from a sharp window function in momenta to the exponential window function used here.}
Notice that the methods used above to compute the $(S/N)$ ratio are not peculiar to the resonant model and can be applied also to other shapes, an example with a different interaction being given in App.~\ref{app:snr_der}. 

From Eq.~\eqref{eq:SNR_final}, it is evident that for $\omega$ somewhat larger than the naive cutoff $4 \pi f$, the signal-to-noise ratio initially grows with $n$, before decaying exponentially. Let us now identify the dominant $n$-point function in this regime. Using the Stirling formula, one gets
\begin{equation}\label{eq:n!asym}
    \frac{1}
    {n! (n-1)! (n-2)!} 
    \simeq
    \left(
    \frac{n}{2 \pi}
    \right)^{3/2}
    \left(
    \frac{\ex{}}{n}    \right)^{3n}\,,
\end{equation}
which is a good approximation for $n \gtrsim 3$, and from which Eq.~\eqref{eq:SNR_finite_3} gives:
\begin{align}\label{snr_n!}
  \bigg(\frac{S}{N}\bigg)^2_n 
	\simeq
    \frac{3}{32 \sqrt{2\pi}} 
    \alpha\, \tilde{b}^2\, n^{3/2}
    \left(\frac{\omega}{4 \pi f}\right)^{-4}
    \bigg[\frac{\omega}{4 \pi f} \left(\frac{{\ex{}}}{n}\right)^{3/2}  \bigg]^{2n}
	N_{\rm modes}
	\;.
\end{align}
Taking the derivative of Eq.~\eqref{snr_n!} with respect to $n$, one finds that the signal is maximum for
\begin{align}\label{N_max_N!}
    n_{\rm max} 
    = 
    \left(
    \frac{\omega}{4 \pi f}
    \right)^{2/3}
    \;,
\end{align}
where we notice that, remarkably, this coincides with the value that maximizes the $n \to n$ scattering amplitude, reported in Eq.~\eqref{eq:n_max_M}.
Plugging \eqref{N_max_N!} into \eqref{snr_n!}, we thus obtain
\begin{align}\label{eq:snr_max_n!}
  \bigg(\frac{S}{N}\bigg)^2 \bigg|_{n_{\rm max}} 
    = 
    \frac{3 \b^2}{16 \sqrt{2} P_\zeta^{1/4} }
    \left(\frac{\omega}{4 \pi f}\right)^{-5/2}
    \exp{
    \left[
    3 
    \left(\frac{\omega}{4 \pi f}\right)^{2/3}
    \right]
    }
    N_{\rm modes}
	\;,
\end{align}
where we have replaced $\alpha$ by $P_\zeta^{-1/4}(\omega/f)^{1/2}$.
We checked that terms subleading in $\alpha$ in the correlators that we have neglected (i.e.~the second term in Eq.~\eqref{eq:col_B_N}) give small corrections at the level of the signal-to-noise ratio.\footnote{
These corrections could become large given the presence of the large parameter $n_{\rm max}$. 
In deriving the correlators at large $\alpha$, Eq.~\eqref{eq:corre_res}, one uses a saddle-point approximation since the integrands are peaked at the saddle point, $-k_{\rm t}\eta_s \sim (-k \eta_s) n = \alpha$ (assuming that all momenta are of the same order). As we increase $n$, this early-time saddle point shifts to later times and we might worry that the flat-space limit used in the mode function becomes unreliable.    
Notice that, in deriving this saddle point, one uses that the wave modes for $\zeta$ behave as $(1 - \ii k_i \eta) \exp(\ii k_i \eta) \sim -\ii k_i \eta \exp(\ii k_i \eta)$ at early times. When we compute the $n$-point function, we need to multiply $n$ of these wave modes. 
Keeping the first correction in the product and assuming all $k_i \sim k$ are of the same order, one obtains corrections $\sim n / (-k \eta)$.
Therefore, when evaluated at $n = n_{\rm max}$ of Eq.~\eqref{N_max_N!} this goes as
$n_{\rm max} / (- k \eta_s) \simeq n_{\rm max}^2 / \alpha$.
For the values of $\alpha$ we are interested in, this quantity is small, at most of order $\sim 1 / 10$, and can be safely neglected (see the values entering in Figs.~\ref{fig:snr_vs_n} and \ref{fig:snr_contour}). 
In any case, these small corrections in principle can be included in the calculation of the signal-to-noise ratio.
} 

\section{Results}\label{sec:result}

\subsection{Measurable $n$-point functions}

When interpreting the result \eqref{eq:snr_max_n!}, one should not forget that the maximum value that $\omega/4 \pi f$ can take depends on the amplitude of the oscillations $\tilde{b}$. The smaller $\tilde{b}$, the larger the window $4\pi f < \omega < \Lambda_{\rm cutoff}$, but the latter only grows logarithmically, approximately following Eq.~\eqref{eq:cutoff_resonant}.
Considering a background frequency $\omega$ at the cutoff $\Lambda_{\rm cutoff}$, Eq.~\eqref{eq:cutoff_resonant} gives the corresponding value of $\tilde{b}$ as a function of $\Lambda_\textrm{cutoff}/(4 \pi f)$:
\begin{align}\label{eq:bt_nmax}
    \tilde{b} 
    \sim
    \left(\frac{\Lambda_\textrm{cutoff}}{4 \pi f}\right)^{-2}
    \exp{
    \left[
    -3 
    \left(\frac{\Lambda_\textrm{cutoff}}{4 \pi f}\right)^{2/3}
    \right]}
    = 
    \frac{1}{n_{\rm max}^3} \ex{-3 n_{\rm max}}
    \;,
\end{align}
where in the last equality we have used Eq.~\eqref{N_max_N!} to express $\tilde{b}$ as a function of the corresponding $n_{\rm max}$. This relation, valid in the limit of large $n_{\rm max}$, shows that $\b$ must be taken exponentially small in this regime. Notice also that subleading terms in Eq.~\eqref{eq:cutoff_resonant} that we neglected affect the numerical proportionality constant in Eq.~\eqref{eq:bt_nmax}.
Plugging \eqref{eq:bt_nmax} into Eq.~\eqref{eq:snr_max_n!} we obtain the parametric dependence of the $(S/N)$ ratio at its maximum:
\begin{align}
    \bigg(\frac{S}{N}\bigg)^2\bigg|_{n_{\rm max}} 
        \sim
        \frac{
        \ex{-3n_{\rm max}}
        }
        {
        P_\zeta^{1/4}n_{\rm max}^{39/4}} 
        N_{\rm modes} 
        \;,
\end{align}
where we have omitted a factor of order unity.
We see that the $(S/N)$ decreases exponentially as $\tilde{b}$ decreases, or equivalently as $n_{\rm max}$ increases. However, the numerical prefactor is large and allow for a sizeable $(S/N)$ ratio when $n_{\rm max}$ takes intermediate values. We indeed have $P_{\zeta}^{-1/4} \simeq 60$ and, more importantly, the number of modes $N_{\rm modes}$ takes very large values in current and upcoming surveys of large-scale-structure, up to $10^{9}$, see e.g.~\cite{Achucarro:2022qrl}.  

Despite having a $(S/N)$ that peaks for a certain $n_{\rm max}$, there is a large number of $n$-point functions contributing to the signal. To see this we estimate the width of the $(S/N)$ in Eq.~\eqref{eq:SNR_final} as a function of $n$, in the limit of large $n_{\rm max}$, which we have shown to hold already for $n \gtrsim 3$.
To do so, we write the approximate expression \eqref{snr_n!} as an exponential $(S/N)^2_n = \exp(\Phi(n))$. Then, we expand the exponent $\Phi(n)$ around $n = n_{\rm max}$ to second order in $n - n_{\rm max}$.
Indeed, what controls the width is the second derivative of $\Phi(n)$ with respect to $n$ evaluated at $n = n_{\rm max}$. 
It is straightforward to obtain $\Phi''(n_{\rm max}) \simeq -3/n_{\rm max}.$ 
Therefore, the signal can be approximated around its peak by a Gaussian of width $\sigma^2 = - 1 / \Phi''(n_{\rm max}) \simeq n_{\rm max} / 3$.
This result indicates that the total signal cannot be simply approximated by a single $n$-point function but rather it is given by $\sim n^{1/2}_{\rm max}$ correlators, all contributing similarly.
The total signal-to-noise \eqref{total-SNR}, added in quadrature (and corresponding to the optimal estimator), scales then as $(S/N)^2_{\rm tot} \simeq n^{1/2}_{\rm max} \, (S/N)^2_{\rm max} $ proving an additional (small) boost to the signal.

\begin{figure*}[t!]
	\centering
	 \includegraphics[width=0.6\textwidth]{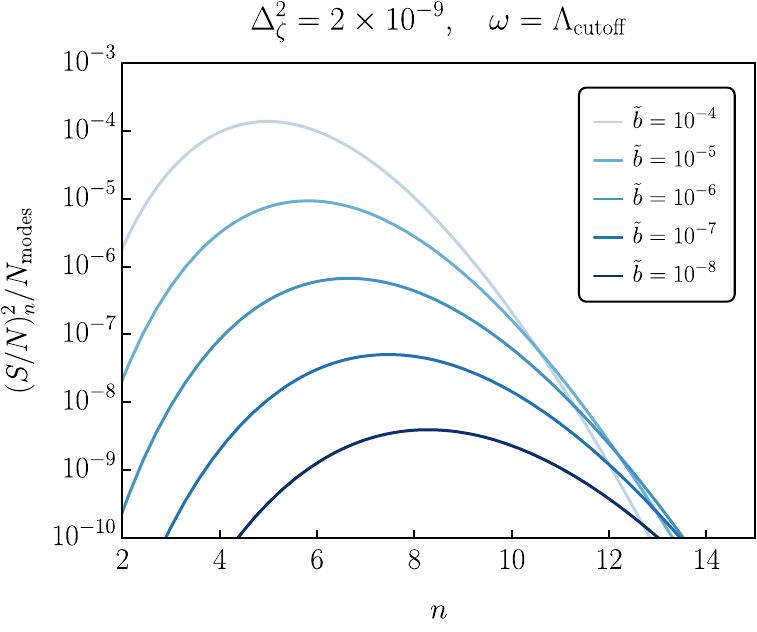}
 	\caption{\;\;Signal-to-noise $(S/N)^2_n$, Eq.~\eqref{eq:SNR_final}, over the number of modes $N_{\rm modes}$, as a function of $n$, evaluated at cutoff frequencies $\omega = \Lambda_{\rm cutoff}$ for different values of $\b$. 
    The cutoff is obtained numerically as explained in the text.}
 	\label{fig:snr_vs_n}
\end{figure*}

The features of the $(S/N)$ ratio discussed above analytically in the limit of large $n$ are confirmed in Fig.~\ref{fig:snr_vs_n} where we numerically evaluate the exact expression for $({ S / N})^2_n / N_{\rm modes}$, Eq.~\eqref{eq:SNR_final}, at the cutoff, as a function of $n$ and for different values of $\b$. 
To obtain the cutoff without approximation, we first analytically find the maximum of the amplitude $|\Cc M_{n \to n}|$ (Eq.~\eqref{eq:M_n_to_n}). This takes into account corrections to the expression \eqref{eq:n_max_M} that become relevant when $n_{\rm max}$ is not sufficiently large. Next, we evaluate the amplitude at this maximum and use the condition $|\Cc M_{n \to n}|_{n_{\rm max}} = 1$ to determine the cutoff. Explicitly, we use this equation to numerically solve for $\omega / (4 \pi f)$ in terms of $\b$.
Finally, we evaluate $(S/N)_n$ in Eq.~\eqref{eq:SNR_final} at this cutoff; therefore, we obtain an expression for $(S/N)^2_n / N_{\rm modes}$ depending only on $\b$ and $n$.

\begin{figure*}[t!]
	\centering
	 \includegraphics[width=0.6\textwidth]{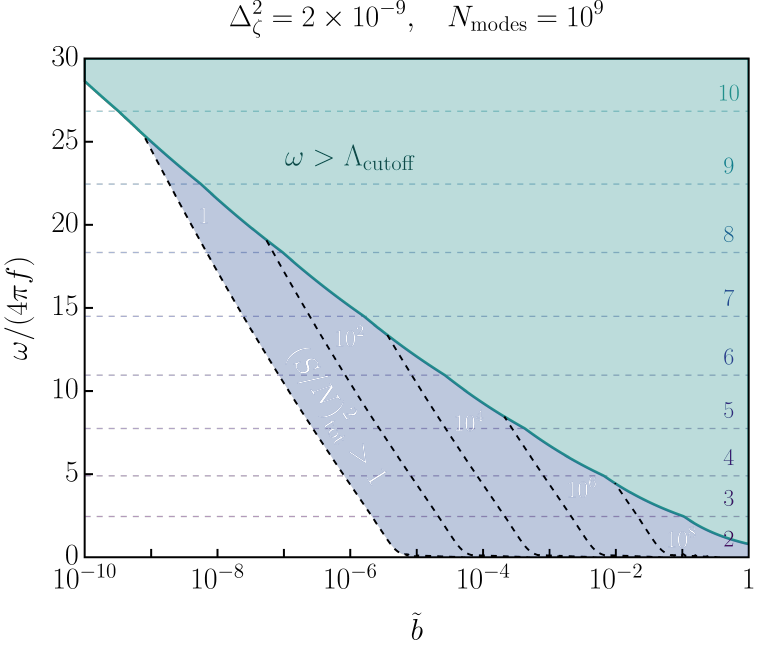}
 	\caption{\;\;Regions of detectability for the resonant non-Gaussianity correlators in the parameter space $(\b$, $\omega)$. 
 	The blue region corresponds to detectable total signal-to-noise ratio in Eq.~\eqref{total-SNR} (for $N_{\rm modes} = 10^{9}$) and frequencies below the cutoff ($|\Cc M_{n \to n}| < 1$ for all $n$'s). Black dashed lines labelled from $1$ to $10^{8}$ indicate the values of ${(S/N)}_{\rm tot}^2$. 
 	The horizontal dashed lines, labelled by the numbers $2, 3,\ldots 10$, denote which correlation function dominates the $(S/N)$ ratio. For instance, in region 3 we have $(S/N)^2_{ n = 3} > (S/N)^2_{n =2}$. 
 	In the green region, the frequency $\omega$ is above the cutoff.}
 	\label{fig:snr_contour}
\end{figure*}

In Fig.~\ref{fig:snr_contour}, we evaluate the signal-to-noise ratio at the cutoff and study the detectability of the resonant signal by scanning the full two-dimensional parameter space $(\b, \, \omega)$ describing the amplitude and frequency of the oscillations in the potential. We fixed $N_\textrm{modes}=10^9$ as representative of upcoming surveys of large-scale-structure. We see that for $\omega/(4 \pi f) \gtrsim 2.5$, there is more information in non-Gaussianities than in the oscillations of the power spectrum. More generally, this results uncover a large region of parameter space where perturbative control is guaranteed ($\omega < \Lambda_\textrm{cutoff}$) and $n$-point functions with $3 \lesssim n \lesssim 9$ are potentially detectable, corresponding to the range $350 \lesssim \alpha \lesssim 1000$.\footnote{Notice that this range of $\alpha$ is obtained using $\omega/f = \alpha^2 P_\zeta^{1/2}$ and $P_\zeta = 4 \pi^2 \Delta_\zeta^2$ with $\Delta_\zeta^2 = 2\times 10^{-9}$ by CMB observations.}

\subsection{Reaching high frequency}
To enter the regime of interest of this paper, one needs fast oscillations $\omega > 4 \pi f$ corresponding to $\alpha > 210$.
This is a quite high frequency, but that can be constrained by current and future surveys. 
Let us analyze this point, concentrating for simplicity on the analysis of the power spectrum described by the following expression:
\begin{equation}
	\frac{k^3}{2 \pi^2}\braket{\zeta_{\vect k} \zeta_{-\vect k}}'
	= 
	\Delta_\zeta^2
	\left[
		1 
		+ 
		A_\textrm{log} \cos (\alpha \log k / k_{\star} + \varphi)
	\right]
	\;,
\end{equation} 
where we neglect the tilt $n_s$ for simplicity, $k_{\star} = 0.05 \, {\rm Mpc}^{-1} / h$ is the pivot scale, $A_\textrm{log}$ and $\alpha$ are, respectively, the amplitude and frequency of the resonant feature and $\varphi$ is a phase. 
Comparing this parameterisation with the expression \eqref{eq:corre_res}, we see that $A_\textrm{log}$ is related to $\alpha$ and $\b$ as $A_\textrm{log} = \sqrt{\alpha \pi/2}\, \b$.

In order to observe the oscillation of a signal, one needs at least few data points within a single oscillation (Nyquist-Shannon frequency).
The resolution in Fourier space is limited by the size of the survey $L$.
For the model at hand, the frequency of oscillations in Fourier space is not fixed but depends logarithmically on the scale. In particular, around a given point $k$, the frequency of oscillation is of order $\sim \alpha / k$. We want this frequency to be smaller than the limit discussed above: $\alpha / k < L$. This is easier to achieve if one looks at the shortest modes in the survey, so that one gets $ \alpha / (k_{\rm max} L) < 1$.

Current observational bounds are limited to regions where $0 \lesssim \alpha \lesssim 100$ and $A_\textrm{log} \lesssim 0.04$ (see for instance \cite{Antony:2024vrx,Planck:2018jri}). Exploring higher values for $\alpha$ requires surveys with larger values of $k_{\rm max} L$. For the current bounds, $\alpha \simeq 100$ corresponds to $\omega / (4 \pi f) \simeq 0.2$ (taking $\Delta_\zeta^2 \simeq 2 \times 10^{-9}$).  
In this regime, we are at frequencies much lower than the naive cutoff $4\pi f$ and, as confirmed in Fig.~\ref{fig:snr_contour}, the signal is dominated by the correction to the power spectrum. 
Upcoming and future surveys will reach $\Delta k \simeq 0.001 \, h \, {\rm Mpc}^{-1}$, and possibly higher $k_{\rm max}$, which allows to reach $\alpha \simeq 360$ \cite{Calderon:2025xod, Mergulhao:2023ukp}. This would allow to explore the region $\omega / (4 \pi f) \simeq 3$, where the bispectrum starts to dominate the signal.

\section{Conclusions and future directions}\label{sec:conclusions}

In this paper we identified a new regime of resonant non-Gaussianity in which the impact on the power spectrum is negligibly small, while higher-point correlation functions can be measurable. 

A natural question that arises is how to experimentally probe such a signal, given that a brute-force analysis of an $n$-point function for large $n$ appears unfeasible. It will be interesting to explore whether other approaches, not directly based on $n$-point functions (simulation-based inference and field-level  inference:  see for instance \cite{Schmidt:2025iwa} and references therein) are better tools to constrain exotic initial statistics. 

The signal also features very fast oscillations, presenting a significant experimental challenge. 
In the standard case ($\omega \lesssim 4\pi f$), when the oscillations of the power spectrum are the most relevant feature, the maximum frequency is $\alpha \simeq 210$: this is an important target in the search of oscillations in the power spectrum. Our new regime features even faster oscillations, up to $\alpha \sim 1000$.

From a theoretical perspective, it is important to systematically explore models that produce large, potentially measurable $n$-point functions. Some such models have already been proposed, typically based on multi-field dynamics (see, e.g.~\cite{Chen:2018uul, Panagopoulos:2019ail,Munchmeyer:2019wlh}). One promising direction is to investigate localized features to determine if there are regimes with a peculiar hierarchy of correlators, as in the case studied here. In this context, the concept of ``boundaries" in field space, introduced in \cite{Cheung:2024wme}, is worth further exploration.

While we did not attempt to derive an explicit UV completion of the theory in this paper, we adopted the unitarity limit as a bound. This approach may be overly permissive, and explicit UV completions could impose additional constraints. For instance, the models studied in \cite{Hook:2023pba} suggest a parametrically smaller cutoff. An explicit UV completion could also introduce interesting phenomenological features, see e.g.~\cite{Chen:2022vzh,Qin:2023ejc,Werth:2023pfl,Pinol:2023oux,Pajer:2024ckd} for recent related analyses.

Another question that arises is the relationship between the perturbativity of the theory and its non-Gaussianity. Naively, the two concepts are closely related: a theory becomes strongly coupled when it becomes fully non-Gaussian, i.e.~when the free theory is no longer a good approximation. However, for the model under consideration, these concepts diverge slightly. Indeed, as one moves along the line that describes ``strong coupling" in Fig.~\ref{fig:snr_contour}, the theory remains cosmologically rather Gaussian, at least for small enough $\tilde b$. The simple interaction $\dot\pi^n$ studied in App.~\ref{app:snr_der} can be a good starting point for investigating this phenomenon.

\section*{Acknowledgements}
It is a pleasure to thank I.~Antoniadis, G.~A.~Palma, R.~Rattazzi, L.~Senatore, S.~Sypsas, G.~Villadoro and Z.~Z. Xianyu for useful discussions.
V.~Y. is supported by World Premier International Research Center Initiative (WPI), MEXT, Japan.
The work of V.~Y. is supported by grants for development of new faculty staff, Ratchadaphiseksomphot Fund, Chulalongkorn University and by the NSRF via the Program Management Unit for Human Resources \& Institutional Development, Research and Innovation Grant No.~B41G680029. We thank the feedback of participants to the program \href{https://indico.ijclab.in2p3.fr/event/11373/}{CoBALt} held at the Institut Pascal at Universit\'e Paris-Saclay with the support of the program “Investissements d’avenir” ANR-11-IDEX-0003-01. 

\appendix
\section{S-matrix calculation in flat space}\label{app:flat_scattering}
Following \cite{Hook:2023pba}, in this appendix we compute the $n \rightarrow n$ matrix element $\Cc M_{n \rightarrow n}$ in flat space. 

\subsection*{Phase-space integral}
Before proceeding to the calculation, let us first evaluate the phase-space integral that usually appears in scattering-amplitude computations, see also \cite{Chang:2019vez}. We define 
\begin{align}\label{eq:phase_space_n}
	\Omega_n 
	&\equiv
	\int \left[\prod_{i = 1}^n \frac{\de^3 \vect p_i}{(2\pi)^3 2 E_{\vect{p}_i}}\right]
	\, 
	(2\pi)^4 \delta^{(4)}\big(P^\mu - \sum_{j = 1}^n p_j^\mu\big)  
\;,
\end{align}
where $n$ is the number of particles in the final state, $p^\mu_i$ and $E_{\vect{p}_i}$ are the four-momentum and the energy of particle $i$, and $P^\mu$ is the total four-momentum of the final state. Then, replacing the delta function in the above expression with exponentials gives	
\begin{align}
	\Omega_n 
	&= 
	\int \de^4 x 
	\int 
	\left[\prod_{i = 1}^n \frac{\de^3 \vect p_i}{(2\pi)^3 2 E_{\vect{p}_i}}\right]
	\ex{-\ii P \cdot x} 
	\ex{+\ii \sum_i^n p_i \cdot x} 
	\\ 
	&= 
	\int \de^4 x 
	\, 
	\ex{-\ii P \cdot x} 
	\bigg[ 
		\int  \frac{\de^3 \vect p_i}{(2\pi)^3 2 E_{\vect{p}_i}} 
		\, 
		\ex{\ii p_i \cdot x}
	\bigg]^n 
	\;, 
	\label{eq:omega_N_D}
\end{align}
where the symbol dot here refers to the four dimensional inner product with Minkowski metric. 
Note that, for massive particles, we have $E_{\vect{p}} = \sqrt{\vect{p}^2 + m^2}$. 
We notice that the integral in the square bracket is nothing but the Wightman function $D(x)$ for a massive scalar in four dimensions, defined as
\begin{align}\label{eq:wightman_fn}
	D(x) 
	&\equiv 
	\int 
	\frac{\de^3 \vect p_i}{(2\pi)^3 2 E_{\vect{p}_i}} 
	\, 
	\ex{\ii p_i \cdot x}  
	= 
	\frac{m}{(2\pi)^2 \sqrt{x^2}} K_1(m \sqrt{x^2}) 
	\;,
\end{align} 
where in the second equality we have assumed that $x^2 > 0$ (spacelike separation) and $K_1$ denotes the modified Bessel function of the second kind (see e.g.~Ch.~5.2 of \cite{Weinberg:1995mt}). Note that for timelike separation ($x^2 < 0$), one instead obtains $D(x) = i m H_1^{(2)}(m \tau)/(8 \pi \tau)$, where $\tau$ is a proper time and $H_1^{(2)}$ is the Hankel function of the second kind. In the massless limit, which is the one relevant for us, the expression \eqref{eq:wightman_fn} becomes
\begin{align}\label{eq:wightman_massless}
	D(x) = \frac{1}{(2\pi)^2 [\vect{x}^2 - (t - \ii \varepsilon)^2]} \;,
\end{align}
where we have properly added the $\ii \varepsilon$ prescription for the un-ordered (Wightman) two-point function. 
Then, using \eqref{eq:wightman_massless} in \eqref{eq:omega_N_D} we have
\begin{align}
	\Omega_n 
	&= 
	\frac{1}{(2 \pi)^{2n}} 
	\int \de^4 x 
	\, 
	\frac{\ex{-\ii  P \cdot x}}
	{\left[\vect{x}^2 - (t - \ii \varepsilon)^2\right]^n} 
	\\
	&=
	\frac{1}{(2 \pi)^{2n}} 
	\int_{-\infty}^{+\infty} \de t \, 
	\ex{\ii  P^0 t}
	\int_{0}^{+\infty} \de r \, 
	\frac{4 \pi r^2}{\left[r^2 - (t - \ii \varepsilon)^2\right]^n} 
	\label{eq:omega_n_euclidean}
	\;,
\end{align}
where in the second line we used $P^\mu = (P^0, \vect{0})$. The integral over $r$ can be performed in closed form, with the $\ii\varepsilon$ providing the prescription to avoid the singularities.
We have
\begin{equation}
	\int_{0}^{+\infty} \de r \, 
	\frac{4 \pi r^2}{\left[r^2 - (t - \ii \varepsilon)^2\right]^n}
	= 
	-\ii
	(-1)^n
	\frac{\pi^{3/2} \, \Gamma(n - 3/2)}{\Gamma(n)} (t - \ii \varepsilon)^{3 - 2n}
	\;.
	\label{eq:dr_int_Omega}
\end{equation}
The final integral over $t$ can be performed using Cauchy's theorem by noticing the presence of a pole at $t = \ii \varepsilon$. Moreover, since $P^0 > 0$ the integrand decays exponentially in the region ${\rm Im\, }{(t)} > 0$. 
We therefore deform the contour to a patch $\mathcal C_{\varepsilon}$ that encloses the pole and write $t = \ii \varepsilon + z$ around the pole: 
\begin{align}
	\int_{-\infty}^{+\infty} \de t \, 
	\ex{\ii  P^0 t}
	(t - \ii \varepsilon)^{3 - 2n}
	&= 
	\ex{- P^0 \varepsilon}
	\oint_{\mathcal C_{\varepsilon}}
       \de z\,
	\ex{\ii P^0 z}
	z^{3 - 2 n}
	\\
	&=
	\sum_{k = 0}^{\infty}
	\ex{- P^0 \varepsilon}
	\frac{(\ii P^0)^{k}}{k!}
	\oint_{\mathcal C_{\varepsilon}}
	\de z \,
	z^{3 - 2n + k}
	\\
	&=
	\ex{- P^0 \varepsilon}
	\frac{(\ii P^0)^{2n - 4}}{(2 n - 4)!}
	2 \pi \ii 
	\;.
	\label{eq:dt_int_Omega}
\end{align}
In the last line we selected the simple pole, i.e.~$k = 2 n - 4$.
Combining Eqs.~\eqref{eq:dr_int_Omega} and \eqref{eq:dt_int_Omega} in \eqref{eq:omega_n_euclidean} and sending $\varepsilon \to 0$ we finally obtain
\begin{equation}
	\Omega_n 
	= 
	\frac{(-1)^n}{(2 \pi)^{2n - 1}}
	\frac{\pi^{3/2} \, \Gamma(n - 3/2)}{\Gamma(n) (2 n - 4)!} 
	(\ii P^0)^{2n - 4}
	\;.
\end{equation}
Writing the energy $E \equiv P^0$ we can simplify the result to
\begin{align}\label{eq:omega_n}
	\Omega_n 
	= 
	\frac{1}{8\pi} 
	\left(\frac{E}{4\pi}\right)^{2n - 4} 
	\frac{1}{(n-1)! (n-2)!} 
	\;,
\end{align}
where we used the property $\Gamma(z) \Gamma(z + 1/2) = 2^{1-2z}\sqrt{\pi}\,\Gamma(2z)$ with $z \in \mathbb{C}$.
This result agrees for instance with \cite{Kleiss:1985gy}, even though it is obtained in an independent manner.

\subsection*{Scattering amplitude of normalisable states}
Plane-wave states are delta-function normalisable and therefore are not suitable for being used in evaluating the perturbative unitarity bound.  Instead of using standard partial-wave amplitudes, one can consider the more convenient states:
\begin{align}\label{eq:normalized_state}
	\ket{P, n}
	= C \int \de^4 x\, 
	\ex{\ii P \cdot x} \big[\phi^{(-)}(x) \big]^n \ket{0} \;,
\end{align}
where $C$ is a normalisation factor, $P^\mu$ is the total four-momentum of the state, $\phi^{(-)}(x)$ contains the creation operator $a_{\vect{p}}^\dagger$ and $n$ is a positive integer. These correspond to $n$-particle states that are normalisable:
\begin{align}\label{eq:othogonal_state}
	\braket{P', n' | P, n} 
	= 
	(2\pi)^4 \delta^{(4)}(P - P') \delta_{n, n'} \;.
\end{align}
As pointed out in \cite{Chang:2019vez}, the states \eqref{eq:normalized_state} are defined such that they have the largest overlap with the interaction we will consider below. 
We can write the scalar field in terms of creation and annihilation operators as usual
\begin{align}\label{eq:phi_fourier}
	\phi(x) 
	= 
	\int \frac{\de^3 \vect p}{\sqrt{(2\pi)^{3} 2 E_{\vect{p}}}} 
	\bigg[
		a_{\vect{p}}\,\ex{\ii p \cdot x} + a^\dagger_{\vect{p}}\,\ex{-\ii p \cdot x} 
	\bigg] 
	= 
	\phi^{(+)}(x) + \phi^{(-)}(x) 
	\;,
\end{align}
with $[a_{\vect{p}}, a^\dagger_{\vect{q}}] = \delta^{(3)}(\vect{p} - \vect{q})$. 

Now let us impose the normalisation condition \eqref{eq:othogonal_state} to fix the coefficient $C$ in \eqref{eq:normalized_state}.
From Eq.~\eqref{eq:normalized_state} we find that 
\begin{align}
	\braket{P', n' | P, n} 
	&= 
	|C|^2 \int \de^4 x \de^4 y \, 
    \ex{\ii P \cdot x} \ex{-\ii P' \cdot y}  
    \braket{0 | \big[\phi^{(+)}(y) \big]^{n'} \big[\phi^{(-)}(x) \big]^n | 0} \;, \\
	&= 
	|C|^2 
	\int 
	\bigg[\prod_{i = 1}^{n'} \de \tilde{p}_i  \bigg]
	\bigg[\prod_{j = 1}^{n}  \de \tilde{q}_j  \bigg]
	\int \de^4 x \de^4 y \, 
	\ex{\ii (P - \sum_i {q}_i)\cdot x}
	\ex{-\ii (P' - \sum_j {p}_j)\cdot y} 
	\times \nonumber \\ 
	&\hspace{0.5cm} 
	\times 
	\braket{0 | 
	\bigg[\prod_{i=1}^{n'} a_{\vect{p}_i}\bigg] 
	\bigg[\prod_{j = 1}^{n}a_{\vect{q}_j}^\dagger\bigg] | 0} 
	\\
	&= 
	|C|^2 \delta_{nn'} 
	\int 
	\bigg[\prod_{i = 1}^{n} \de \tilde{p}_i  \bigg]
	\bigg[\prod_{j = 1}^{n}  \de \tilde{q}_j  \bigg]
	\int \de^4 x \de^4 y \, 
	\ex{\ii (P - \sum_i {q}_i)\cdot x} 
	\ex{-\ii(P' - \sum_j {p}_j)\cdot y} 
	\times \nonumber \\ 
	&\hspace{0.5cm} \times 
	\bigg( \prod_{i=1}^{n}  
	\delta^{(3)}(\vect{p}_i - \vect{q}_i) + {\rm perm.} \bigg) 
	\;,
\end{align}
where $\de \tilde{p}_i \equiv \de^3 \vect p_i/(\sqrt{(2\pi)^3 2 E_{\vect{p}_i}})$ and in the last line we have used the commutation relation of $a_{\vect{p}}$ and $a_{\vect{p}}^\dagger$.
Then, performing the integrals over $\vect{q}_i$ with the delta functions leads to 
\begin{align}
	\braket{P', n' | P, n} 
	&= 
	|C|^2 
	\delta_{nn'} n! 
	(2 \pi)^4 \delta^{(4)}(P - P') 
	\int 
	\bigg[
		\prod_{i=1}^n\frac{\de^3 p_i}{(2\pi)^3 2 E_{\vect{p}_i}} 
	\bigg] 
	(2\pi)^4 \delta^{(4)}\big(P - \sum_{i=1}^n p_i\big) 
	\\
	&= 
	|C|^2 \delta_{nn'} n! 
	(2 \pi)^4 \delta^{(4)}(P - P') 
	\, \Omega_n \;,
\end{align}
where in the first line we have evaluated the integrals over $x$ and $y$, leading to two delta functions. In the second line we recognized the appearance of Eq.~\eqref{eq:phase_space_n}.
Therefore, using the formulas \eqref{eq:omega_n} and \eqref{eq:othogonal_state}, we obtain 
\begin{align}\label{eq:norm_factor}
|C|^{-2} = n! \, \Omega_n = \frac{1}{8\pi} \bigg(\frac{E}{4\pi}\bigg)^{2n-4} \frac{n!}{(n-1)! (n-2)!}  \;.
\end{align}

At this point we are ready to evaluate our scattering amplitude.
We consider the matrix element $\Cc M_{n\rightarrow n}$ due to the interaction
\begin{align}
	\mathcal{L}_{2n} 
	=  \frac{\lambda_{2n} \phi^{2n}}{(2n)!} 
	\;.
\end{align}
As usual, the S-matrix can be written as $S = \mathds{1} + \ii T$. 
Here, we focus on the leading order in $\lambda_{2n}$, which corresponds to the leading-order terms in $\tilde{b}$ of the resonant model discussed in the main text.
Then, we have 
\begin{align}
	\braket{P', n'| \ii T | P, n}  
	&= 
	\braket{P', n' | \ii \int \de^4 x\, \frac{\lambda_{2n} \phi(x)^{2n}}{(2n)!} | P, n} 
	\\
	&= 
	\frac{\ii \lambda_{2n}}{(2n)!} |C|^2 
	\int \de^4 x \, \de^4 y_1 \, \de^4 y_2 
	\, 
	\ex{\ii P \cdot y_2} 
	\ex{-\ii P'\cdot y_1} 
	\braket{0 |  [\phi^{(+)}(y_1)]^{n'} \phi(x)^{2n} [\phi^{(-)}(y_2)]^n | 0} 
	\;, \label{eq:iiT_evalu}
\end{align}
where in the last line we have used Eq.~\eqref{eq:normalized_state}. Using \eqref{eq:phi_fourier} in Eq.~\eqref{eq:iiT_evalu}, we then obtain 
\begin{align}
	\braket{P', n'| \ii T | P, n} 
	&=  
	\frac{\ii \lambda_{2n}}{(2n)!} 
	|C|^2 
	\int 
	\de^4 x \, 
	\de^4 y_1 \, 
	\de^4 y_2 \, 
	\ex{\ii P \cdot y_2} 
	\ex{-\ii P'\cdot y_1} 
	\bigg[
		\prod_{i = 1}^{n'} 
		\de \tilde{p}_i 
		\, \ex{\ii p_i \cdot y_1}
	\bigg] 
	\bigg[
		\prod_{j = 1}^n \de \tilde{q}_j \, \ex{-\ii q_j \cdot y_2}
	\bigg]  
	\times \nonumber \\
	&\hspace{0.5cm} 
	\times 
	\bigg[
		\prod_{k=1}^{2n} \de \tilde{\ell}_k 
	\bigg] 
	\braket{0 |  
	\bigg[\prod_{i = 1}^{n'} a_{\vect{p}_i} \bigg] 
	\bigg[\prod_{k=1}^{2n}
	\bigg(a_{\vect{\ell}_k} \ex{\ii \ell_k \cdot x} + a^\dagger_{\vect{\ell}_k} \ex{-\ii \ell_k \cdot x} \bigg) 
	\bigg] 
	\bigg[\prod_{j = 1}^n a^\dagger_{\vect{q}_j}\bigg] 
	| 0} 
	\\ 
	&= 
	\frac{\ii \lambda_{2n}}{(2n)!} 
	|C|^2 
	\delta_{n'n} (2n)! 
	\int \de^4 x \de^4 y_1 \de^4 y_2 \, 
	\ex{\ii P \cdot y_2} 
	\ex{-\ii P'\cdot y_1} 
	\bigg[\prod_{i = 1}^n \frac{\de^3 p_i}{(2\pi)^3 2E_{\vect{p}_i}} \, \ex{\ii p_i \cdot (y_1 - x)}\bigg] 
	\times \nonumber \\
	&\hspace{0.5cm} 
	\times 
	\bigg[\prod_{j = 1}^n \frac{\de^3 q_j}{(2\pi)^3 2E_{\vect{q}_j}} \, \ex{-\ii q_j \cdot (y_2 - x)}\bigg] 
	\;,
\end{align}
where in the last equality we have used the contractions among the creation and annihilation operators. Note that this contraction indeed gives rise to the factor $(2n)!$.
Then, integrating over $y_1$, $y_2$ and $x$ yields 
\begin{align}
	\braket{P', n'| \ii T | P, n} 
	&= 
	\ii 
	\lambda_{2n} |C|^2 \delta_{n'n} 
	\int \bigg[\prod_{i = 1}^n \frac{\de^3 p_i}{(2\pi)^3 2E_{\vect{p}_i}} \bigg] 
	(2\pi)^4 \delta^{(4)}\big(P' - \sum_{i=1}^n p_i\big)  
	\times \nonumber \\ 
	& \hspace{0.5cm} \times  
	\bigg[\prod_{j = 1}^n \frac{\de^3 q_j}{(2\pi)^3 2E_{\vect{q}_j}} \bigg]
	(2\pi)^4 \delta^{(4)}\big(P - \sum_{j = 1}^n q_j\big) 
	(2\pi)^4 \delta^{(4)}\big(P - P'\big) 
	\\
	&=
	\ii \lambda_{2n} |C|^2 \delta_{n' n}(\Omega_n)^2 (2\pi)^4 \delta^{(4)}\big(P - P'\big) \;.
\end{align}
Using the formula \eqref{eq:norm_factor} we therefore obtain
\begin{align}\label{eq:amplitude_E_dep_app}
	\Cc M_{n\rightarrow n}  
	= 
	\frac{\lambda_{2n} }{8\pi} \bigg(\frac{E}{4\pi}\bigg)^{2n-4} \frac{1}{n!(n-1)! (n-2)!}  
	\;,
\end{align}
where we have used 
\begin{align}
	\braket{P', n'| \ii T | P, n} 
	= 
	(2\pi)^4 \delta_{n'n}\delta^{(4)}(P - P') \ii \Cc M_{n\rightarrow n} 
	\;.
\end{align}
Indeed, Eq.~\eqref{eq:amplitude_E_dep_app} agrees with the result found in \cite{Chang:2019vez}, which was later used in \cite{Hook:2023pba} (see also Sec.~\ref{sec:cutoff}) to derive the perturbative unitarity bound. 

\section{Degenerate configurations and off-diagonal terms in the covariance matrix}\label{app:off-diagonal}

In our estimators \eqref{eq:estimator_n}, the integral over momenta is restricted to non-degenerate polygons. In this appendix, we show the consequence of taking these configurations into account on off-diagonal terms of the covariance matrix. Interestingly, despite the fact that degenerate polygons are few, we demonstrate that their effect is not volume suppressed, but only parametrically suppressed by the loop-counting parameter of the resonant model.

At the order $\b^2$ at which we are working, the fields $\zeta_{\vect{k_i}}$ in \eqref{eq:estimator_n} are treated as Gaussian, hence $\braket{\hat{\Cc E}_{n} \hat{\Cc E}_{m}}$ automatically vanish for $m = n + (2k + 1)$ with $k = 0,1,2,\cdots$. 
Thus, the non-trivial contributions come from $\braket{\hat{\Cc E}_{n} \hat{\Cc E}_{m}}$ for $m = n + 2k$.
For simplicity, let us first concentrate on $m=n+2$ ($k=1$). In this case, we are forced to contract two $\zeta$'s from $\hat{\Cc E}_{n+2}$, giving a delta function and therefore an overall additional integral compared to the diagonal case \eqref{eq:var_estimator}:
\begin{align}\label{eq:estimator_cov_1}
	\braket{\hat{\Cc E}_{n} \hat{\Cc E}_{n+2}}
	&= \frac{(n+2)!}{2}
	\frac{\Cc V \,}{N_n N_{n+2} n! (n+2)!} 
	\int \prod_{i = 1}^n 
	\bigg[\frac{\de^3 \vect{k}_i}{(2\pi)^3}\bigg] 
	(2 \pi)^{3}
	\delta^{(3)}(\vect k_{\rm t}) 
    \times
    \\
    &
    \hspace{4.5cm} \times
	\int 
	\frac{\de^3 \vect{p}}{(2\pi)^3} 
	\frac{
	\braket{ \zeta_{\vect{k}_1} \cdots \zeta_{\vect{k}_n}}'_{| \b=1}
	\braket{ \zeta_{\vect{k}_1} \cdots \zeta_{\vect{k}_n} \zeta_{\vect{p}} \zeta_{-\vect{p}}}'_{| \b=1}
		}
	{P(k_1) \cdots P(k_n) P(p)}
	\nonumber \\
	&=
	\frac{\Cc V}{2 N_n N_{n+2} n!}
	\frac{2^{n+1} A_n A_{n+2}}{P^{n+1}_{\zeta} \b^2}
	\int \prod_{i = 1}^n 
	\bigg[\frac{\de^3 \vect{k}_i}{(2\pi)^3 k_i}\bigg] 
	(2 \pi)^{3}
	\delta^{(3)}(\vect k_{\rm t}) 
    \times
    \\
    &
    \hspace{4.5cm} \times
	\int 
	\frac{\de^3 \vect{p}}{(2\pi)^3 p} 
	\frac{
		\sin(\alpha \log(k_{\rm t} / k_{\star}))}
	{k_{\rm t}^{n - 3}}
	\frac{
		\sin(\alpha \log((k_{\rm t}  + 2 p)/ k_{\star}))}
	{(k_{\rm t} + 2 p)^{n - 1}}
	\;,
\end{align}
where $(n+2)!/2$ is the number of equivalent contractions and we used the explicit shape of the correlators \eqref{eq:corre_res} in the second line. 
To proceed further, we evaluate the integral over $\vect p$. We have that 
\begin{align}
	\int 
	\frac{\de^3 \vect{p}}{(2\pi)^3 p}
	\frac{
		\sin(\alpha \log((k_{\rm t}  + 2 p)/ k_{\star}))}
	{(k_{\rm t} + 2 p)^{n - 1}}
	&=
	\frac{1}{2 \pi^2}
	\frac{1}{2 \ii}
	\sum_{\sigma=\pm 1}
	\sigma
	\int_{0}^{\infty}
	\de p
	\frac{p }{(k_{\rm t} + 2 p)^{n-1}}
	\left(
		\frac{k_{\rm t} + 2 p}{k_{\star}}
	\right)^{\ii \alpha \sigma}
	\nonumber \\
	&=
	\frac{k_{\rm t}^{3-n}}{8 \pi^2}
	\frac{1}{2 \ii}
	\sum_{\sigma=\pm 1}
	\sigma
	\frac{(k_{\rm t} / k_{\star})^{\ii\alpha \sigma}}{(\ii \alpha \sigma - n + 3)(\ii \alpha \sigma - n + 2)}
	\nonumber \\
	&
	\simeq
	-
	\frac{k_{\rm t}^{3-n}}{\alpha^2 8 \pi^2}
	\sin(\alpha \log (k_{\rm t} / k_{\star}))
	\;.
\end{align}
In the last step, we have taken the limit $\alpha \gg n$ in the denominator inside the summation over $\sigma$, which is well within the regime of interest, see, e.g.~Fig.~\ref{fig:snr_vs_n}. We can insert this approximate result in Eq.~\eqref{eq:estimator_cov_1}. Moreover, we notice that $A_{n+2} = (\alpha^4 P^2_{\zeta} / 4) A_{n}$ and we therefore obtain
\begin{align}
	\braket{\hat{\Cc E}_{n} \hat{\Cc E}_{n+2}}
	&\simeq
	-
	\frac{\Cc V}{2 N_n N_{n+2} \cdot n!}
	\frac{2^{n} \alpha^2 P_{\zeta} A^2_n}{16 \pi^2  P^{n}_{\zeta} \b^2}
	\int \prod_{i = 1}^n 
	\bigg[\frac{\de^3 \vect{k}_i}{(2\pi)^3 k_i}\bigg] 
	(2 \pi)^{3}
	\delta^{(3)}(\vect k_{\rm t}) 
	\frac{
		\sin^2(\alpha \log(k_{\rm t} / k_{\star}))}
	{k_{\rm t}^{2n - 6}}
	\nonumber \\
	&=
    -
    \frac{1}{2}
	\frac{\alpha^2 P_{\zeta}}{16 \pi^2} \frac{N_n}{N_{n+2}}
	\braket{\hat{\Cc E}_{n}^2}
	\;, \label{eq:E_nE_n+2}
\end{align}
where we compare with the expression \eqref{eq:var_estimator} of the variance $\braket{\hat{\Cc E}_{n}^2}$ computed by restricting to non-degenerate configurations. Equivalently, Eq.~\eqref{eq:E_nE_n+2} can be rewritten as
\begin{align}
\frac{|\braket{\hat{\Cc E}_{n} \hat{\Cc E}_{n+2}}|}{\sqrt{\braket{\hat{\Cc E}_{n}^2} \braket{\hat{\Cc E}_{n+2}^2}}}=\frac{1}{2}
	\frac{\alpha^2 P_{\zeta}}{16 \pi^2} \left(\frac{N_n}{N_{n+2}}\right)^{1/2} \;,
    \label{size-off-diagonal}
\end{align}
where we have used $\braket{\hat{\Cc E}_{n}^2} = N_n^{-1}$.
Moreover, using $(S/N)_n^2=\b^2 N_n$ and the explicit expression for $(S/N)_n^2$ (Eq.~\eqref{eq:SNR_final}), one gets
\begin{align}
\left(\frac{N_n}{N_{n+2}}\right)^{1/2}=\frac{1}{(\omega/4 \pi f)^2} (n+2)^{1/2} (n+1) n(n-1)^{1/2}\,.    
\end{align}
Using Eq.~\eqref{N_max_N!}, one sees that in the high frequency regime of interest, this ratio approaches unity when evaluated at $n_\textrm{max}$ where the signal is maximum (this makes sense, as the $(S/N)$ ratio weakly depends on $n$ near the maximum).
More generally, the numerical results in Fig.~\ref{fig:snr_vs_n} show that this ratio never compensates in \eqref{size-off-diagonal} for the large suppression by the loop-counting parameter $\alpha^2 P_{\zeta} / (16 \pi^2)= \omega \Delta_\zeta/(8 \pi f)$, which is at most $6 \cdot 10^{-4}$ in the relevant range $\omega/(4 \pi f) \lesssim 25$, see Fig.~\ref{fig:snr_contour}.\footnote{Notice that different $n$-pt functions can have different phases in general, which is neglected in Eq.~\eqref{eq:corre_res}. Including them would make the result even more suppressed.} Similar results can be obtained for other off-diagonal terms $\braket{\hat{\Cc E}_{n} \hat{\Cc E}_{n+2k}}$, which are easily seen to be suppressed by additional loop factors, $(\alpha^2 P_\zeta)^k$.


\section{Signal-to-noise ratio of $\zeta'^n$ interaction}\label{app:snr_der}
Here, we want to derive the $(S/N)$ ratio of the $n$-point function computed from the derivative coupling $\zeta'^n$. 
This will show that the methods used to obtain the $(S/N)$ ratio are not limited to the resonant model.
This self-interaction can be seen to arise in the EFT of inflation \cite{Cheung:2007st}. Let us consider indeed the following operator:
\begin{align}\label{eq:S_int}
S_{\rm int} = \int \de^4 x \sqrt{-g}\, \frac{M_n^4}{n!} (\delta g^{00})^n \;,
\end{align}
where $M_n^4$ is the EFT parameter and $\delta g^{00} \equiv 1 + g^{00}$ is the perturbation of $g^{00}$ around the background metric.
In the decoupling limit, as usual we introduce the St\"uckelberg field $\pi(t,\vect{x})$ through $t \to t + \pi(t,\vect{x})$, so that $\delta g^{00} \rightarrow -2 \dot{\pi} + (\partial_\mu \pi)^2$.
Using the St\"uckelberg trick in the action \eqref{eq:S_int} and neglecting the time-dependence of $M_n^4$, one reads off the interaction $\dot{\pi}^n$, 
\begin{align}
 	S_\pi^{(n)} = \int \de t \de^3 \vect{x}~a^3(t) \frac{M_n^4}{n!} \dot{\pi}^n \;,
\end{align}
where we have omitted terms that contain $(\partial_\mu \pi)^2$ since they have higher powers of the perturbation compared with the one we kept. As usual, introducing $\de t = -\de \eta /(H\eta)$ and $\zeta = -H\pi$, we obtain the action for the curvature perturbation in conformal time
\begin{align}\label{eq:action_zeta}
	S_\zeta 
	= 
	\int \de \eta\, \de^3 \vect{x} 
    \,
	\bigg\{ 
	\frac{1}{2 \eta^2 P_\zeta } 
	\left[\zeta'^2 - (\partial_i \zeta)^2 \right] 
	+ \frac{\lambda_n}{n!}\eta^{n-4} \zeta'^n
	 \bigg\} \;,
\end{align}
where we have included the quadratic part for $\zeta$, prime denotes a derivative with respect to $\eta$, $P_\zeta \equiv H^4/(2M^2_{\rm Pl} |\dot{H}|)$ and $\lambda_n \equiv M_n^4/H^4$. 

Our goal is to derive the tree-level $n$-point function using perturbation theory.
From the interaction contained in Eq.~\eqref{eq:action_zeta}, one uses the \textit{in}-\textit{in} formalism to derive the $n$-point correlation function at first order in $\lambda_n$:
\begin{align}
	\braket{\zeta_{\vect{k}_1} \cdots \zeta_{\vect{k}_n}}'
	&= 
	2^{1-n} 
	\lambda_n P_\zeta^{n} 
	\Gamma(2n-3)
	 k_{\rm t}^{3-2n}
	\bigg(\prod_{i=1}^n k_i\bigg)^{-1} 
	\\ 
	&\equiv 
	\tilde{A}_n \tilde{B}_n(k_i) \;,
	\label{eq:n-pt_fn}
\end{align}
where
\begin{align}
    \tilde{A}_n \equiv 2^{1-n} \lambda_n P_\zeta^{n} \Gamma(2n-3)  \;, \qquad \tilde{B}_n(k_i) = k_{\rm t}^{3-2n}\bigg(\prod_{i=1}^n k_i\bigg)^{-1} \;.
\end{align}
Let us now compute the $(S/N)$ ratio associated with the above $n$-point function. Using the formula \eqref{eq:n-pt_fn} in Eq.~\eqref{individual-SNR}, we have 
\begin{align}
    \bigg(\frac{S}{N}\bigg)^2_n 
    = 
    \frac{2^n \Cc V \tilde{A}_n^2}{n! P_\zeta^n} 
    \frac{1}{\Gamma(4n -6)} 
    \int \prod_{i = 1}^n \bigg[\frac{k_i \de^3 \vect{k}_i}{(2\pi)^3}\bigg] 
    \int \de^3 \vect{x} \, 
    \ex{\ii \vect{x}\cdot\sum_i\vect{k}_i} 
    \int_{-\infty}^0 
    \de \tau \, 
    (-\tau)^{4n-7} \ex{k_{\rm t}\tau}  
    \;, 
    \label{eq:app_SNR_1}
\end{align}
where we have used the formula \eqref{eq:k_T_integral} and the power spectrum $P(k) = P_\zeta/(2k^3)$.
Similar to the calculation in Sec.~\ref{sec:SNR}, here we employ an exponential window function $\ex{-k_i/k_{\rm max}}$ to regularise the momentum integrals.
Then, in Eq.~\eqref{eq:app_SNR_1}, swapping the order of the integrals over $\tau$ and $\vect{k}_i$ gives
\begin{align}
    \bigg(\frac{S}{N}\bigg)^2_n 
    = 
    \frac{2^n \Cc V  \tilde{A}_n^2}{n! \Gamma(4n - 6)P_\zeta^n}  
    \int_{-\infty}^0 \de \tau~(-\tau)^{4n-7} 
    \int \de^3 \vect{x}~
    \bigg[ 
    \int \frac{\de^3 \vect{k}_i}{(2\pi)^3} k_i 
    \, 
    \ex{\ii \vect{x}\cdot\vect{k}_i} 
    \ex{k_{i}\tau} 
    \ex{-k_i/k_{\rm max}}
    \bigg]^n \;,
\label{eq:SNR_zeta'_2}
\end{align}
where we have inserted the exponential window function.
The integral over spatial momenta in the square parenthesis can be performed analytically using spherical coordinates, giving 
\begin{align}
    \int \frac{ \de^3 \vect{k}_i}{(2\pi)^3} 
    k_i
    \, 
    \ex{\ii \vect{x}\cdot\vect{k}_i} 
    \ex{k_{i}\tau} 
    \ex{-k_i/k_{\rm max}} 
    = 
    - \frac{k_{\rm max}^4 [-3 + 6 k_{\rm max}\tau + k_{\rm max}^2( r^2 - 3  \tau^2)]}{
    \pi^2
    [1 - 2 k_{\rm max}\tau +k_{\rm max}^2 (r^2 + \tau^2)]^3} 
    \;,
\end{align}
where $r = |\vect{x}|$. Plugging this into Eq.~\eqref{eq:SNR_zeta'_2} and evaluating the integral over $\vect{x}$ we get
\begin{align}
    \bigg(\frac{S}{N}\bigg)^2_n  
    = 
    \frac{2^{n+2}\Cc V  \tilde{A}_n^2}{P_\zeta^n} 
    \frac{(-1)^n  k_{\rm max}^{4n-3}F(n)}{\pi^{2n-1} n! \Gamma(4n -6)}   
    \int_{-\infty}^0 \de \tau~(-\tau)^{4n-7}   (1 - k_{\rm max} \tau)^{3-4n} 
    \;, 
    \label{eq:SNR_app_3}
\end{align}
where we have defined 
\begin{align}
    F(n) 
    &\equiv 
    \frac{3^{\frac{3}{2} - 2n}}{8} 
    \Gamma(n+1) 
    \bigg\{
    \frac{(-3)^{3n} \sqrt{\pi}}{\Gamma(n + 5/2)} 
    \bigg[
    (2 + n)\,
    {}_2F_1\left(-\frac{1}{2}, 3n; n + \frac{5}{2}; -3\right) 
    - 10\,n\, {}_2F_1\left(\frac{1}{2}, 3n; n+\frac{5}{2}; -3\right)
    \bigg] 
    \nonumber \\ 
    &\quad+ 
    \frac{4\Gamma(2n-3/2)}{\Gamma(3n - 1/2)} 
    {}_2F_1\left(3n, 2n-\frac{3}{2}; 3n-\frac{1}{2}; -\frac{1}{3}\right) 
    \bigg\} 
    \;.
\end{align}
Therefore, performing the integral over $\tau$ in Eq.~\eqref{eq:SNR_app_3} we obtain
\begin{align}
    \bigg(\frac{S}{N}\bigg)^2_n 
    =  
    \frac{48(-1)^n \tilde{A}_n^2}{P_\zeta^n\pi^{2n-3}} 
    \frac{2^nF(n)}{n!\Gamma(4n-3)} 
    N_{\rm modes}
    \;,
\end{align}
where we have replaced $k_{\rm max}^3$ in terms of $N_{\rm modes}$ using Eq.~\eqref{eq:def_N_modes}.

\bibliographystyle{utphys}
\bibliography{biblio}

\end{document}